\documentclass[a4paper,11pt]{article}

\usepackage{contribution}

\newcommand{\weblink}[2][]{%
    \ifthenelse{\equal{#1}{}}%
    {\textnormal{\url{#2}}}%
    {\textnormal{\href{#2}{#1}}}%
}

\newcommand{\acknowledgements}[1]{%
  \bigskip\bigskip
  \textsf{\textbf{\Large Acknowledgements}} \\[2ex]
  {#1}
  \bigskip
}

\def\beq{\begin{equation}}
\def\eeq#1{\label{#1}\end{equation}}
\def\eeqn{\end{equation}}

\def\beqa{\begin{eqnarray}}
\def\eeqa#1{\label{#1}\end{eqnarray}}
\def\eeqan{\end{eqnarray}}

\let\bar=\overbar

\def\Dslash{\not{\hbox{\kern-4pt $D$}}}
\def\dslash{\not{\hbox{\kern-2pt $\del$}}}

\def\msb{{\bar{\ssstyle M \kern -1pt S}}}

\newcommand{\contribution}[7][]{%
  \clearpage
  \thispagestyle{plain}
  \ifthenelse{\equal{#1}{}}
  {\hypersetup{pdftitle={#2}}}
  {\hypersetup{pdftitle={#1}}}
  \hypersetup{pdfauthor={{#3} {#4}}}
  {\centering\normalfont\LARGE\bfseries\sffamily #2 \par\nobreak}
  \lhead{}
  \chead{%
    \textit{\footnotesize XIV International Conference on Hadron Spectroscopy
      (\weblink[\textit{hadron2011}]{http://www.hadron2011.de}), 13-17 June 2011, Munich, Germany}%
  }
  \rhead{}
  \bigskip
  \begin{center}
    {#3} {#4}\ifthenelse{\equal{#6}{}}{}{\footnote{\weblink[#6]{mailto:#6}}}
    \ifthenelse{\equal{#7}{}}{}{#7} \\
    \textit{#5}
  \end{center}
  \bigskip
}

\renewcommand{\abstract}[1]{%
  \begin{center}
    \begin{minipage}{0.85\textwidth}
      \begin{footnotesize}
        #1
      \end{footnotesize}
    \end{minipage}
  \end{center}
  \bigskip
}

\begin{document}

%
%
%
%
%
{  


\newcommand{\MeV}{\ensuremath{\textrm{MeV}}}
\newcommand{\MeVc}{\ensuremath{\textrm{MeV}/c}}
\newcommand{\MeVcc}{\ensuremath{\textrm{MeV}/c^2}}

\newcommand{\GeV}{\ensuremath{\textrm{GeV}}}
\newcommand{\GeVc}{\ensuremath{\textrm{GeV}/c}}
\newcommand{\GeVcc}{\ensuremath{\textrm{GeV}/c^2}}

%

\contribution[Exotic \texorpdfstring{$\eta'\pi^- wave$}{eta-prime pi-
  wave}]  
{The exotic \mathversion{bold}$\eta'\pi^-$ Wave in $190\,\GeV$ $\pi^-p\to\pi^-\eta'p$\mathversion{normal} at
COMPASS}  
{Tobias}{Schl{\"u}ter}  
{Ludwig-Maximilians-Universit{\"at} M{\"u}nchen, Department f{\"u}r Physik, 85748 Garching, GERMANY}
{tobias.schlueter@cern.ch}  
{on behalf of the COMPASS Collaboration}  
%

\abstract{%
  A sample of 35\,000 events of the type
  $\pi^-p\to\eta'\pi^-p_{\textrm{slow}}$ ($\eta'\to\eta\pi^-\pi^+$,
  $\eta\to\gamma\gamma$) with $-t>0.1\,\textrm{GeV}^2/c^2$ was
  selected from COMPASS 2008 data for a partial-wave analysis.  We
  study the broad $P_+$ structure known from previous experiments at
  lower energies, in particular its phase motion relative to the
  $D_+$-wave near the $a_2(1320)$ mass and relative to a broad
  $D_+$-wave structure at higher mass.  We also find the $a_4(2040)$.
  We compare kinematic plots for the $\eta'\pi^-$ and $\eta\pi^-$
  final states.}
%

\section{Introduction}

The existence of strongly-bound, resonant states with quantum numbers
not allowed for a fermion-antifermion system has long been expected.
In the light-quark sector isospin symmetry also disallows such a state
for charged $u\bar d$, $d\bar u$ mesons.  On the other hand, a system
decaying into the two pseudo-scalar mesons $\pi$, $\eta$ ($\pi$,
$\eta'$) with orbital angular momentum $L=1$ has quantum numbers
$J^{PG}=1^{-+}$, and thus any resonant contribution to such a system
would have to be identified with a non-$q\bar q$ resonance.

Indeed, several collaborations claimed the observation of such a
resonance in the $\eta\pi$ system at $m(\eta\pi)\approx
1.4\,\GeVcc$~\cite{Beladidze:1993km,Thompson:1997bs,Abele:1998gn}.
Likewise a very strong $P$-wave contribution was observed in the
$\eta'\pi^-$ system~\cite{Beladidze:1993km,Ivanov:2001rv} at
$m(\eta'\pi)\approx 1.6\,\GeVcc$, but the resonant character of the
structure observed in both systems has been
questioned~\cite{Szczepaniak:2003vg}.

We report on the current status of the analysis of the $\eta'\pi^-$ and
$\eta\pi^-$ final states as observed in the data from the 2008 run of
the COMPASS experiment at CERN and compare our results to the
above-quoted publications.  The COMPASS experimental
setup~\cite{Abbon:2007pq,Alexeev:2011} is a two-stage magnetic
spectrometer attached to the SPS accelerator facility at CERN.  During
the bulk of the 2008/09 campaigns the experiment's goal was the
study of the light meson spectrum.  A $190\,\GeVc$ secondary pion beam
impinged on a liquid hydrogen target.  The main trigger components
were a beam definition trigger together with a recoil proton detector
(RPD).  The RPD ensured an unscathed proton emitted at large angles
with momentum corresponding to momentum transfer $-t \gtrsim
0.1\,\GeV^2$.  A veto counter near the spectrometer entry further
suppressed events with particles emitted at large angles, especially
from target fragmentation~\cite{Schluter:2009,Schluter:2011}.  An
additional veto suppressed events where the beam particle passed
through the target undeflected.  Both stages of the spectrometer are
equipped with tracking detectors and particle identification and
neutral detection by means of electromagnetic and hadronic
calorimetry.  In addition, the first spectrometer stage is equipped
with a Ring-Imaging Cherenkov detector, allowing for particle
identifcation.

\section{Data Selection}
\label{sec:data-selection}

Besides trigger requirements, the events considered were selected by
the following topological criteria: a well-defined primary interaction
vertex inside the target with three outgoing tracks (assumed to be
pions) attached and two clusters in the electromagnetic calorimeters.
In order to select the intended $\pi^-\eta'$ and $\pi^-\eta$ final
states, the invariant mass obtained by attaching the pair of
calorimeter clusters to the primary vertex was required to fall into
the range of the $\eta$ or $\pi^0$ mass, respectively.  The
so-identified neutral particle was then --- after a 1C kinematic fit
--- combined with both possible $\pi^-\pi^+$ pairs.  If the invariant
mass of either combination was found to match the
$\eta'\to\pi^-\pi^+\eta$ or, respectively, $\eta\to\pi^-\pi^+\pi^0$
hypothesis, the event was accepted after additional cuts on the total
momentum.  The procedure yielded 35\,000 events for the $\pi^-\eta'$
final state and 110\,000 events for the $\pi^-\eta$ final state.  The
intermediate steps are illustrated in Fig.~\ref{fig:selection-etap}
for the selection of the $\eta'\pi^-$ final state.  The $\eta\pi^-$
selection is similar.

\begin{figure}[htb]
  \begin{center}
    \includegraphics[width=0.32\textwidth]{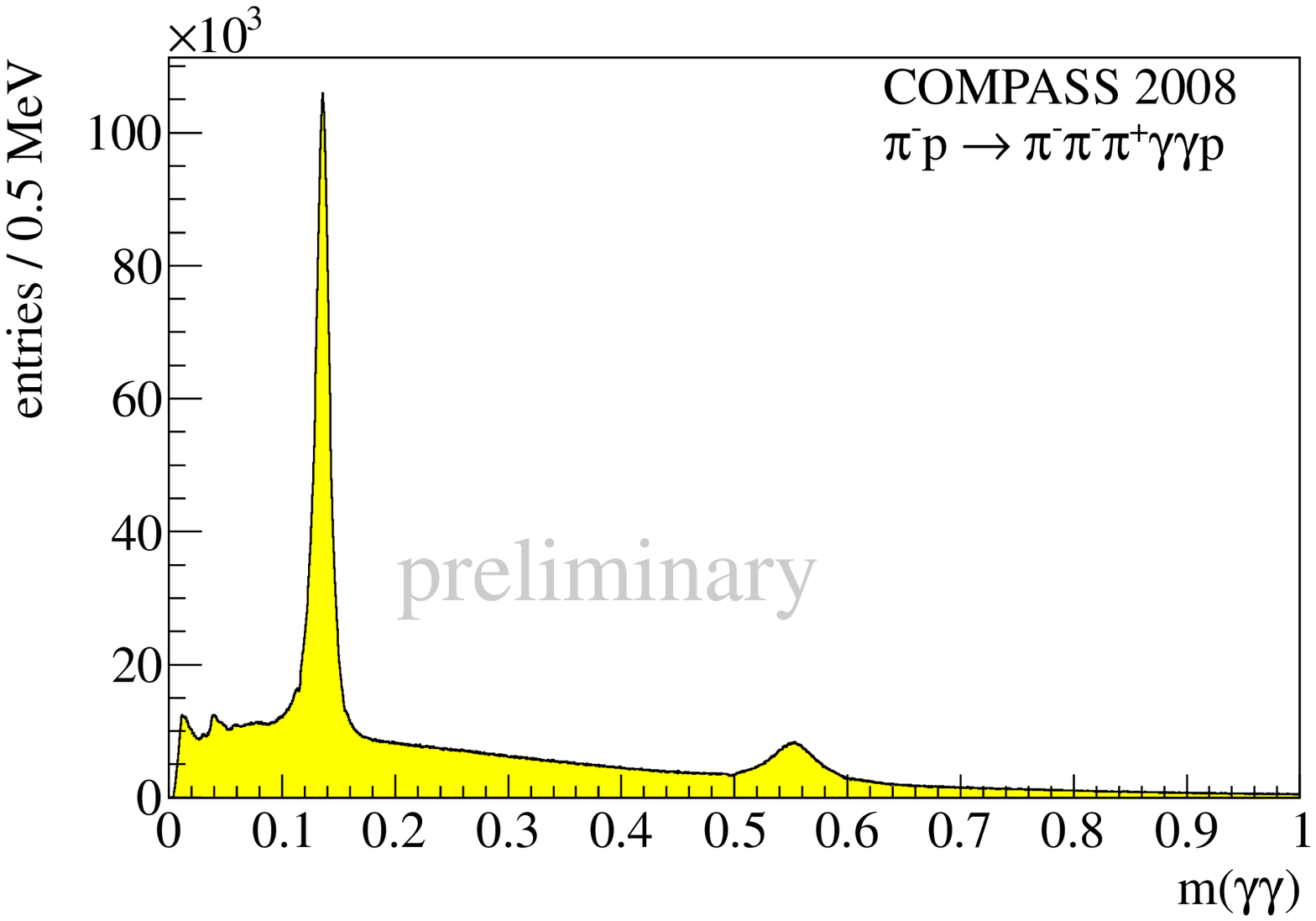}
    \includegraphics[width=0.32\textwidth]{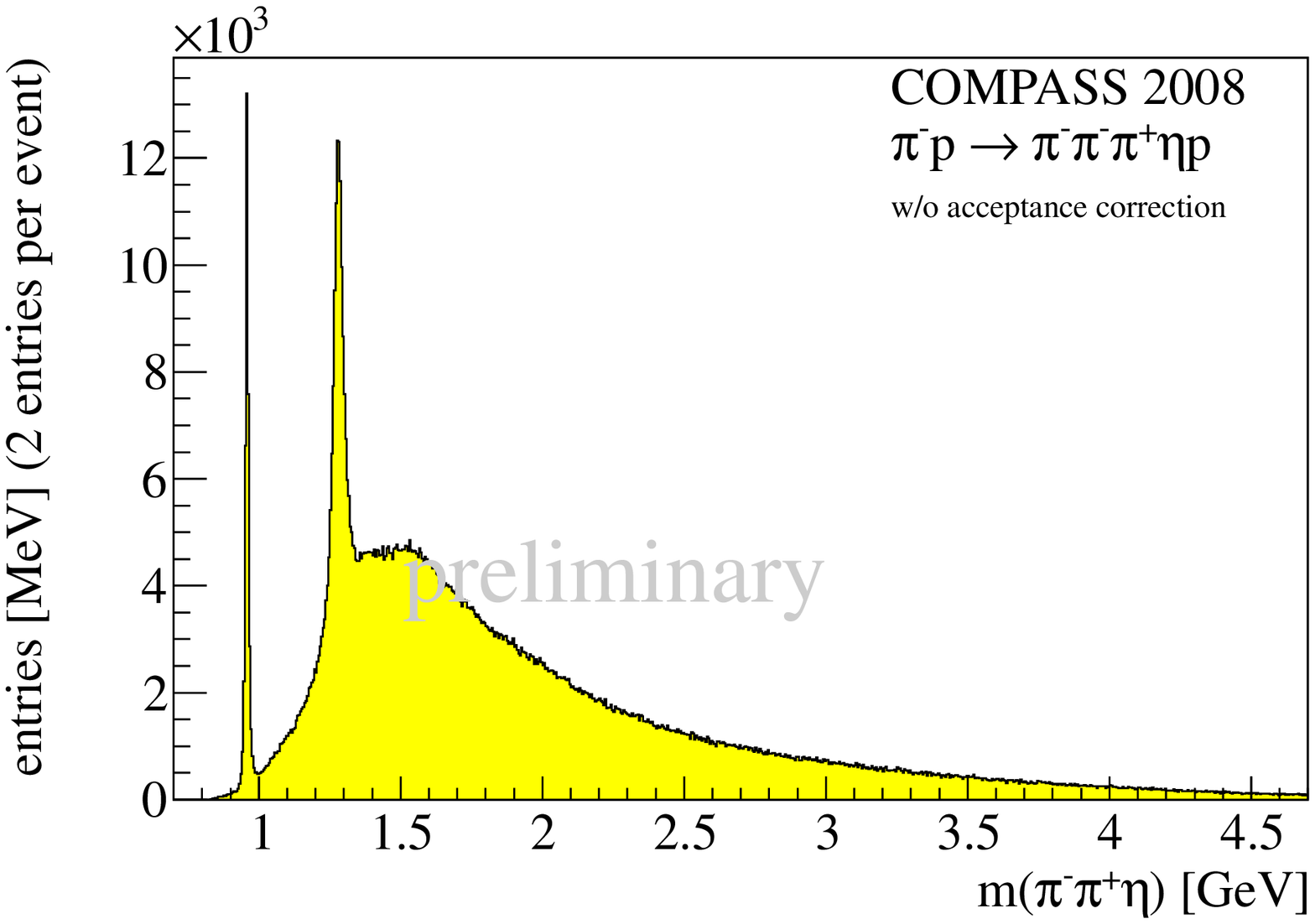}
    \includegraphics[width=0.32\textwidth]{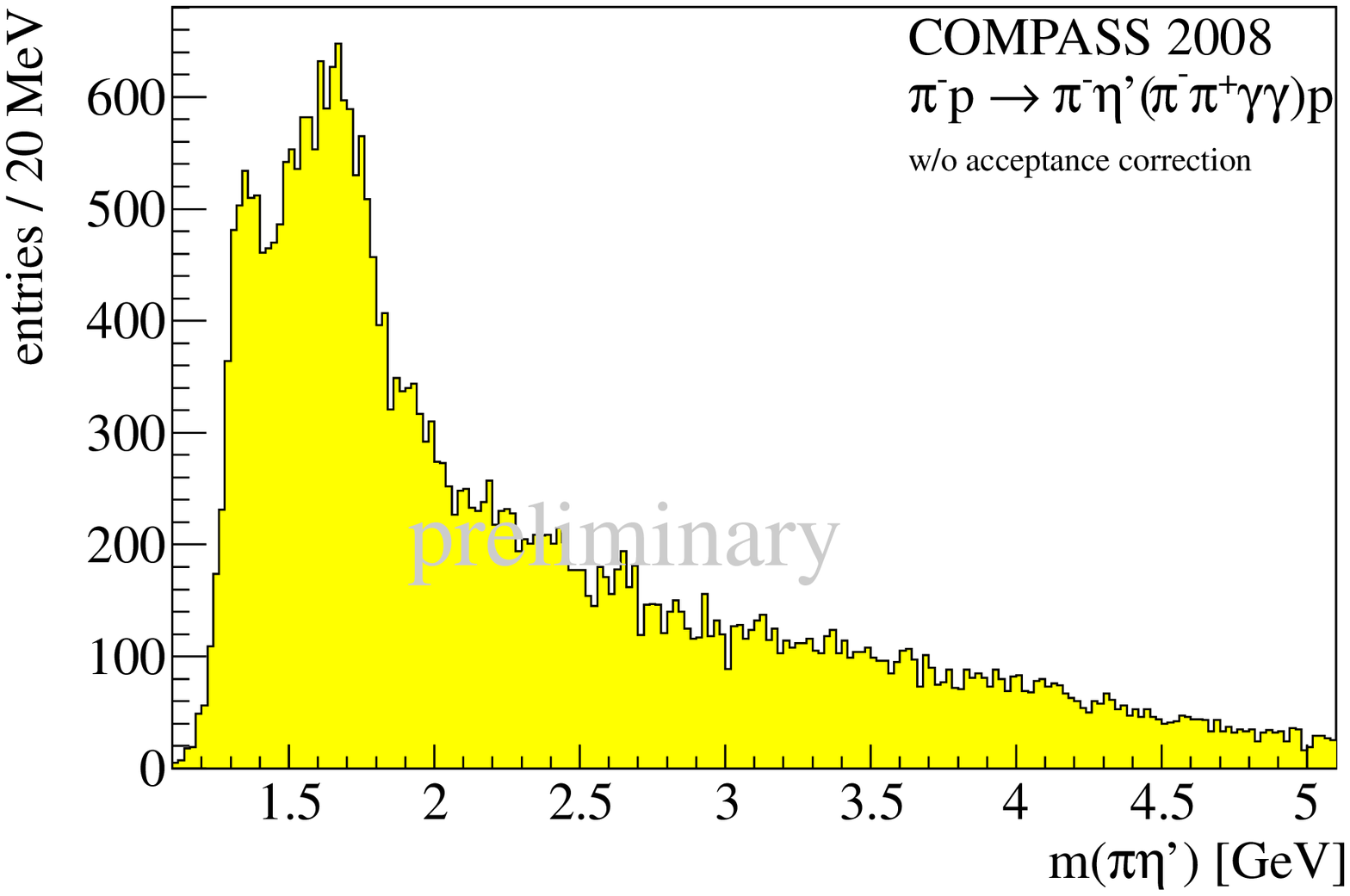}
    \caption{Data selection.  Left: $\gamma\gamma$ mass spectrum for
      events with three charged tracks.  The $\pi^0(135)$ and
      $\eta(548)$ peaks are clearly visible.  The structures below the
      $\pi^0$ mass peak are artefacts of low energetic photon
      reconstruction due to secondary interactions in the detector
      material and to cuts in the reconstruction algorithm. They
      should not be mistaken for any physical signal.  Center:
      $\pi^-\pi^+\eta$ mass spectrum for kinematically complete
      events.  Two peaks corresponding to $\eta'(958)$ and $f_1(1285)$
      stand out.  Right: Final $\pi^-\eta'$ mass spectrum.  The peak
      of the $a_2(1320)$ is visible near threshold.}
    \label{fig:selection-etap}
  \end{center}
\end{figure}

\section{Final-State Kinematics}
\label{sec:final-state-kinem}

The data are expected to be dominated by natural parity exchange waves
with $M=1$, both from the favored pomeron exchange and from the results
of previous analyses.  This is easily verified by plotting the angle
$\phi$ between decay and production plane in the Gottfried-Jackson
frame~\cite{Gottfried:1964nx} where the $\sin^2\phi$ contribution
dominates (not shown).  The remainder of the kinematical information
is contained in the momentum transfer $t$, the polar angle of the
$\eta$ or $\eta'$ meson in the GJ frame $\cos\theta_{\textrm{GJ}}$,
and the invariant mass $m$ of the two-body system under consideration.
In Fig.~\ref{fig:costh-vs-m} the distribution of
$\cos\theta_{\textrm{GJ}}$ as function of $m$ is shown for both the
$\eta\pi^-$ and $\eta'\pi^-$ systems.  Outstanding features are the
occurence of the $a_2(1320)$ meson, a structure near $2\,\GeVcc$
corresponding to the $a_4(2040)$ meson, whose intereference with a
spin-2 background can be made out in the $\eta\pi^-$ data, and a
strong forward-backward peaking for masses above $2\,\GeVcc$, pointing
towards non-resonant contributions.  Especially in the $\eta'\pi^-$
data a strong forward backward asymmetry is observed in the data, with
a fast turnover around the $a_2(1320)$ mass range, corresponding to relative
phase motion of the odd ($P_+$) and even ($D_+$) contributions.

\begin{figure}[htb]
  \begin{center}
    \includegraphics[angle=0,width=0.46\textwidth]{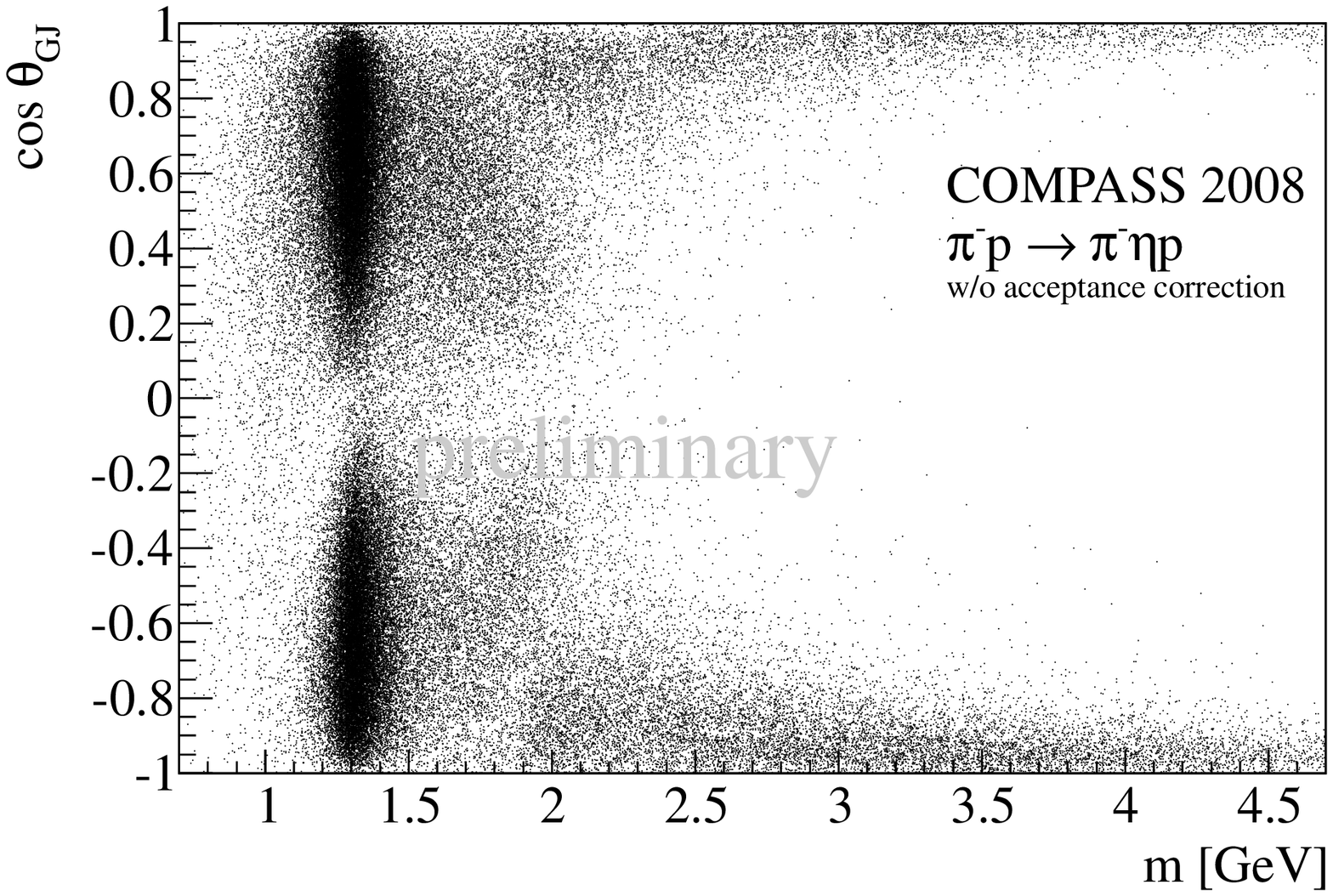}
    \includegraphics[angle=0,width=0.46\textwidth]{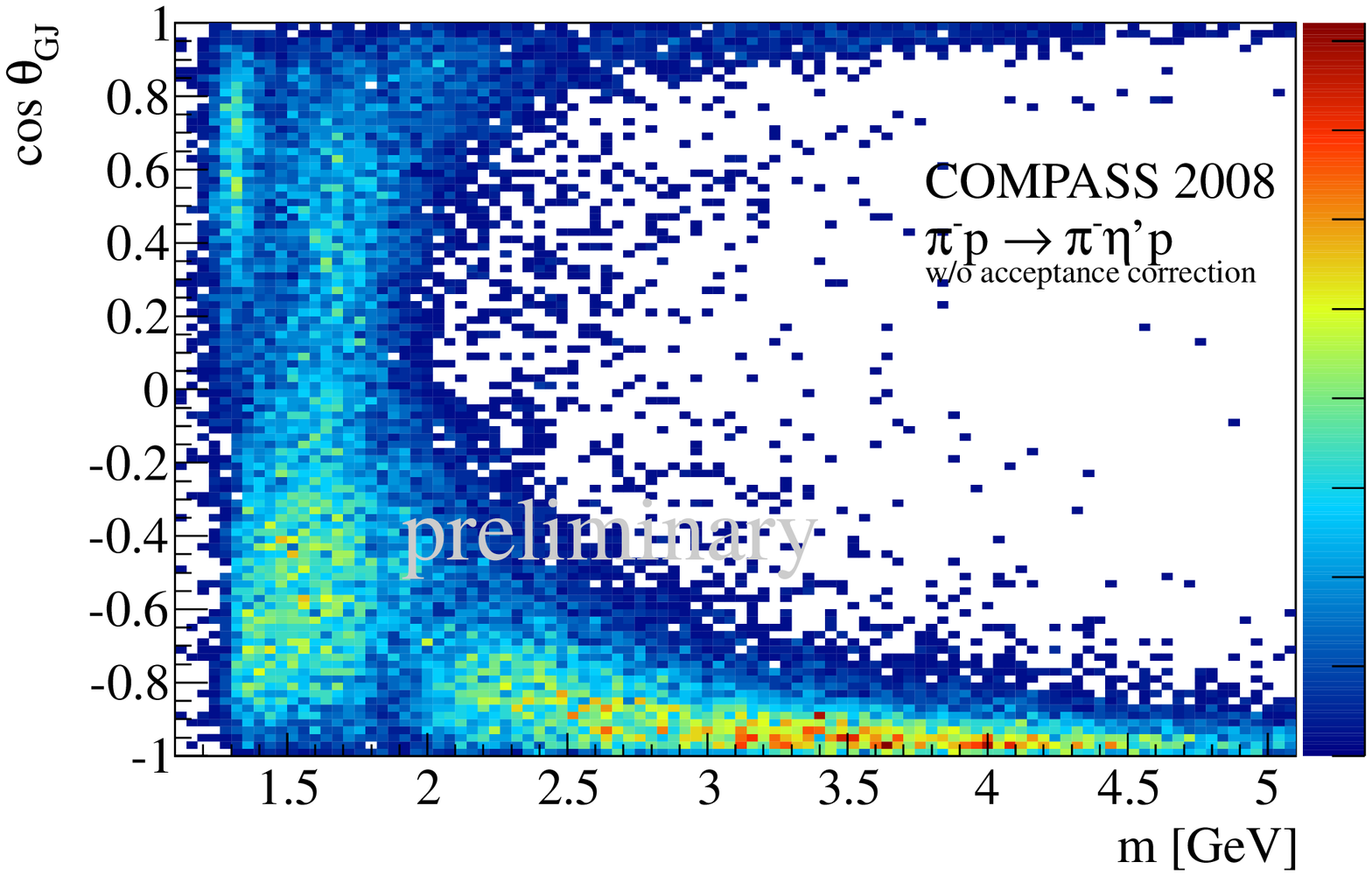}
    \caption{Kinematics.  The distribution in
      $\cos\theta_{\textrm{GJ}}$ as function of invariant mass $m$ (not
      acceptance corrected).  Left for the $\eta\pi^-$ system, right
      for the $\eta'\pi^-$ system.}
    \label{fig:costh-vs-m}
  \end{center}
\end{figure}

The E852 experiment claimed an unusual momentum transfer distribution
in the production of the $\eta'\pi^-$ system~\cite{Ivanov:2001rv}.  We
cannot confirm this observation, see Tab.~\ref{tab:t-distrib}.

\begin{table}[tb]
  \begin{center}
    \begin{tabular}{c|cc}
      mass bin [\GeVcc] & fit with $A\exp(-B|t|)$ &
      fit with $A |t|\exp(-B|t|)$ \\
      \hline
      $m<1.5$ & 5.5 & 8.2 \\
      $1.5 < m < 1.9$ & 5.1 & 7.5 \\
      $1.9 < m < 2.2$ & 4.8 & 7.1 \\
      $2.2 < m < 3$ & 4.6 & 6.9
    \end{tabular}
    \caption{Fit to the slope parameter $B$ in units of $\GeV^{-2}$
      for momentum transfer as function of mass.  E852 found the much
      broader $B=2.93\,\GeV^{-2}$ when fitting with a single
      exponential.}
    \label{tab:t-distrib}
  \end{center}
\end{table}

\section{Partial-Wave Analysis of the \mathversion{bold}\texorpdfstring{$\eta'\pi^-$}{eta'pi-}\mathversion{normal} System}
\label{sec:part-wave-analys}

In this section we present the results of a partial-wave analysis in
mass bins of the $\eta'\pi^-$ data.  The analysis follows the lines of
the previous analyses, allowing $S$, $P$, and $D$ waves with $M\le 1$
in both natural and unnatural exchange.  Additionally, the spin-4,
$M=1$ $G_+$-wave was allowed.  The $\eta'$ was separated from the
$\pi^-\pi^+\eta$ background by introducing its experimental shape as a
pseudo-isobar and fitting in the complete 4-body phase-space where
besides the $\pi^-\eta'$ waves an additional incoherent flat,
phase-space-like contribution was allowed.  From these ingredients an
extended $\log$-likelihood function is constructed and maximized,
taking into account the detector acceptance via normalization
integrals calculated from Monte-Carlo data.  The results for the
intensities of the positive-reflectivity wave are shown in
Fig.~\ref{fig:intensities}.  The relative phases are shown in
Fig.~\ref{fig:phases}.

We confirm the presence of a broad structure in the $P_+$-wave of the
$\eta'\pi^-$.  A fit to the intensity of this structure by a
single-channel relativistic Breit-Wigner shape underestimates the
high-mass side of the distribution.  We confirm the presence of the
$a_4(2040)$ resonance in this channel previously seen by
E852~\cite{Ivanov:2001rv}.  This resonance is also visible in the
$3\pi$ data~\cite{Nerling:2011}.  A resonant interpretation of the
$P_+$-wave would have to be reconciled with the relative phase-motions
compared to the other waves, and the apparent onset of double-Regge
production or similar processes above $\approx 2\,\GeV$ whose low-mass
impact is not yet understood.

\begin{figure}[htb]
  \begin{center}
    \includegraphics[width=0.3\textwidth]{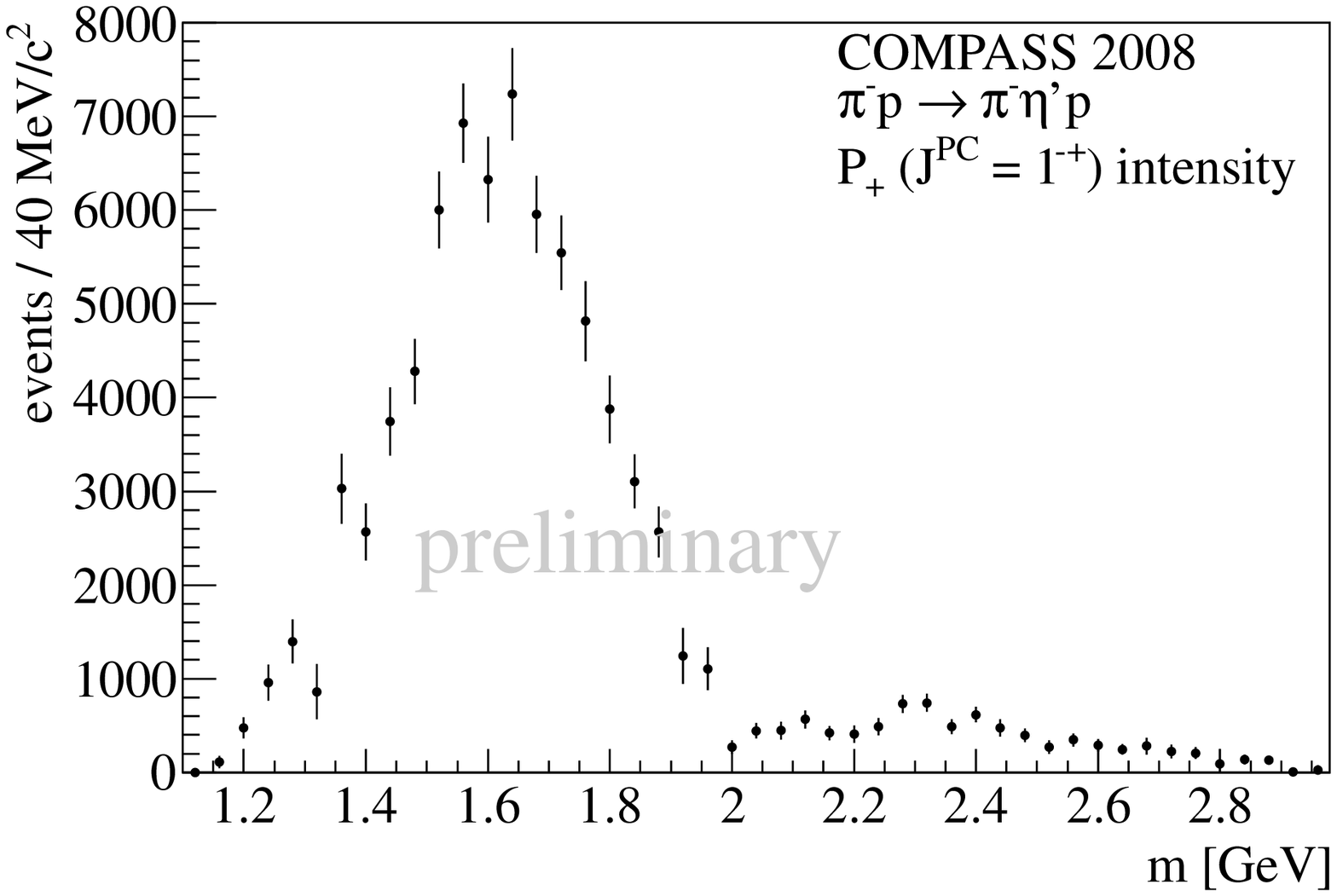}
    \includegraphics[width=0.3\textwidth]{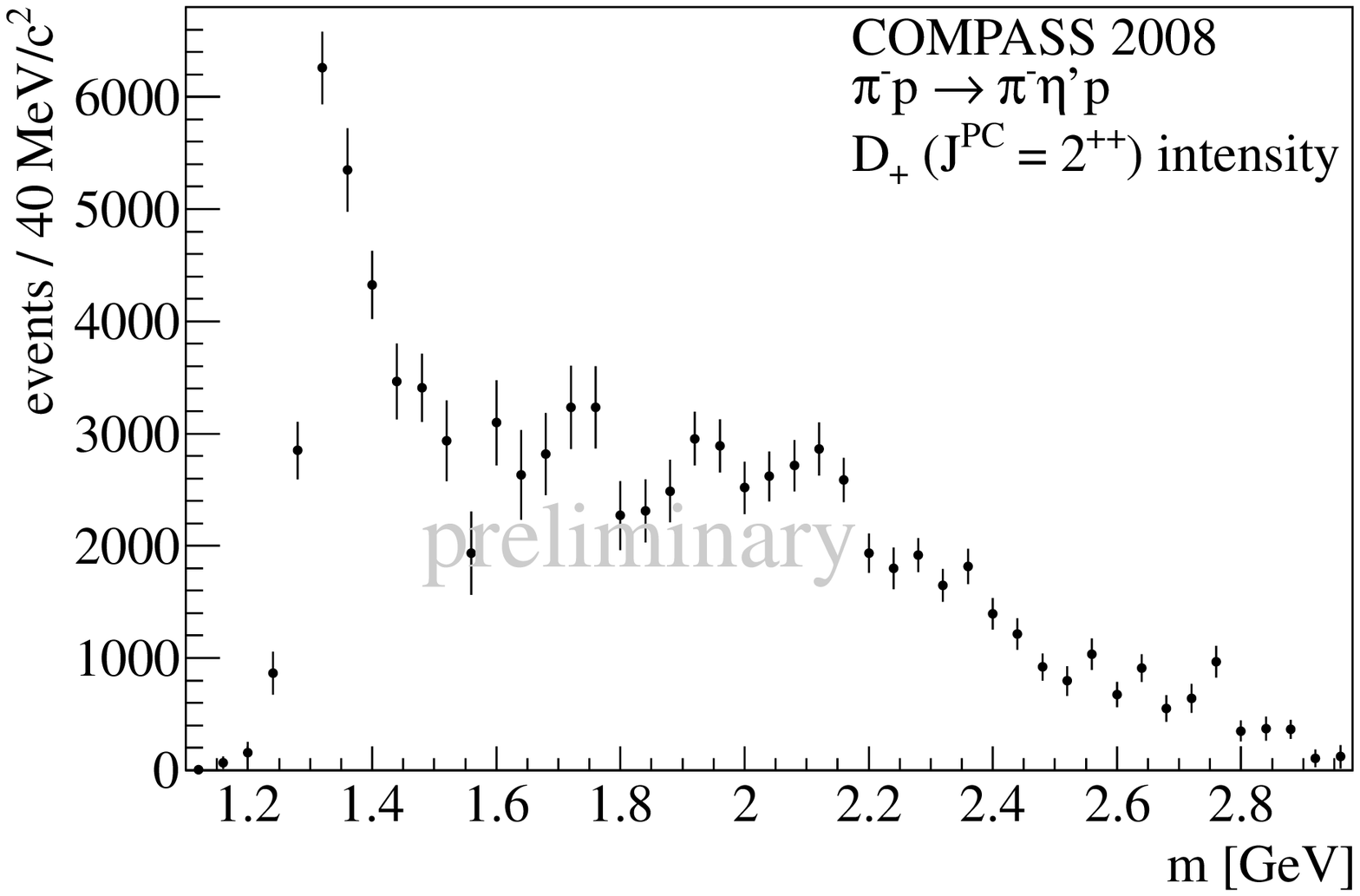}
    \includegraphics[width=0.3\textwidth]{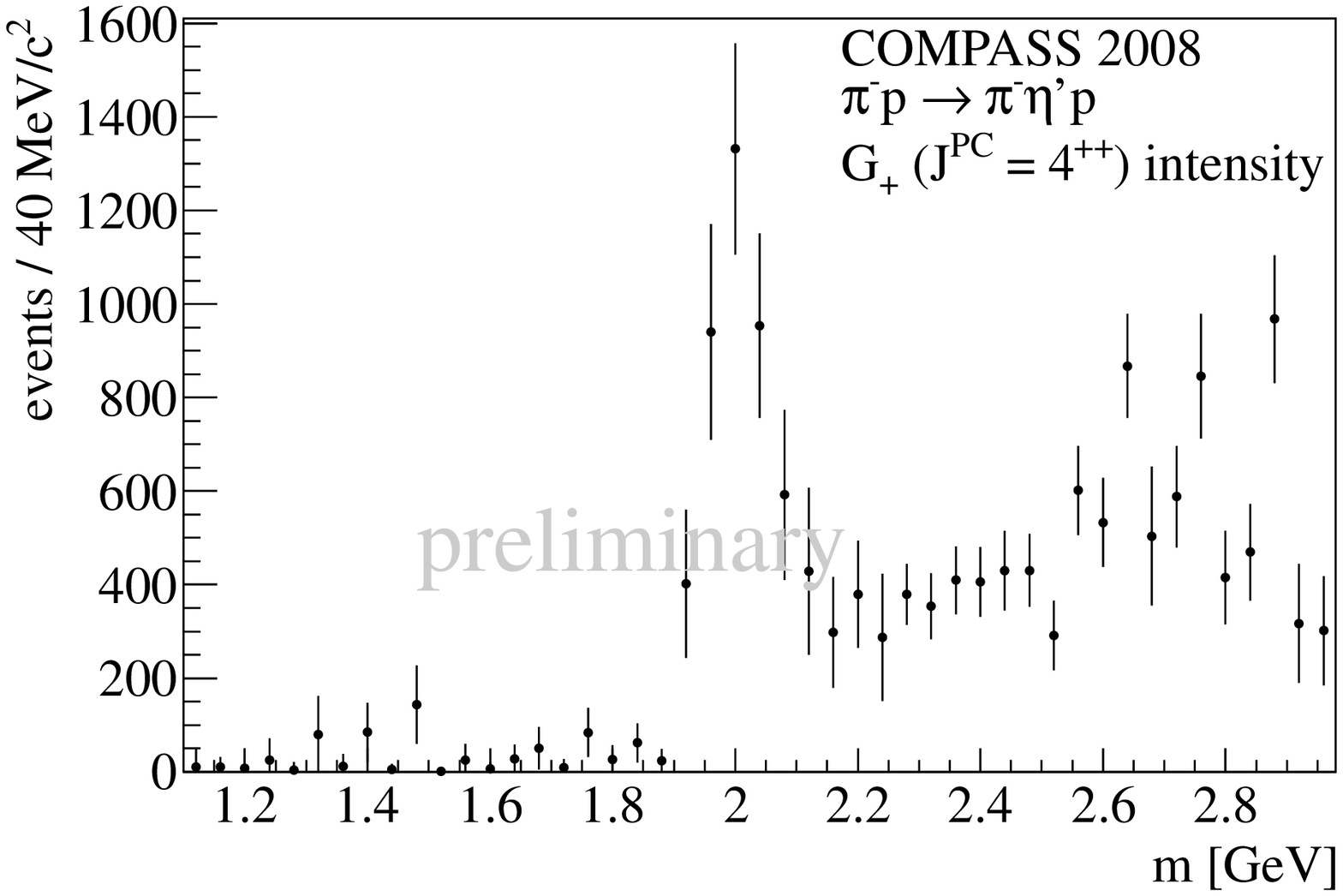}
    \caption{Intensities of the positive-reflectivity waves.  From
      left to right: $P_+$, $D_+$, $G_+$.}
    \label{fig:intensities}
  \end{center}
\end{figure}

\begin{figure}[htb]
  \begin{center}
    \includegraphics[width=0.3\textwidth]{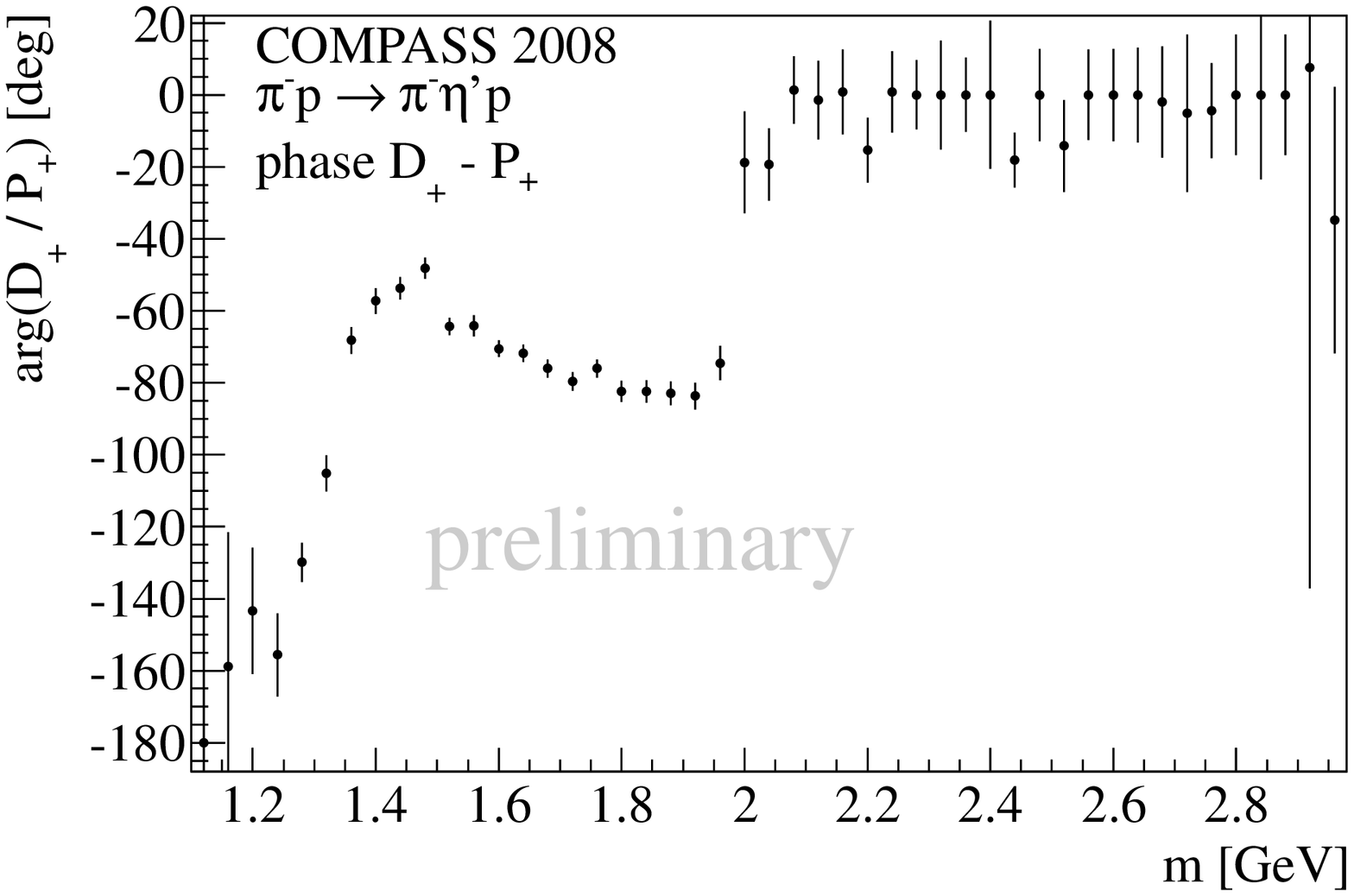}
    \includegraphics[width=0.3\textwidth]{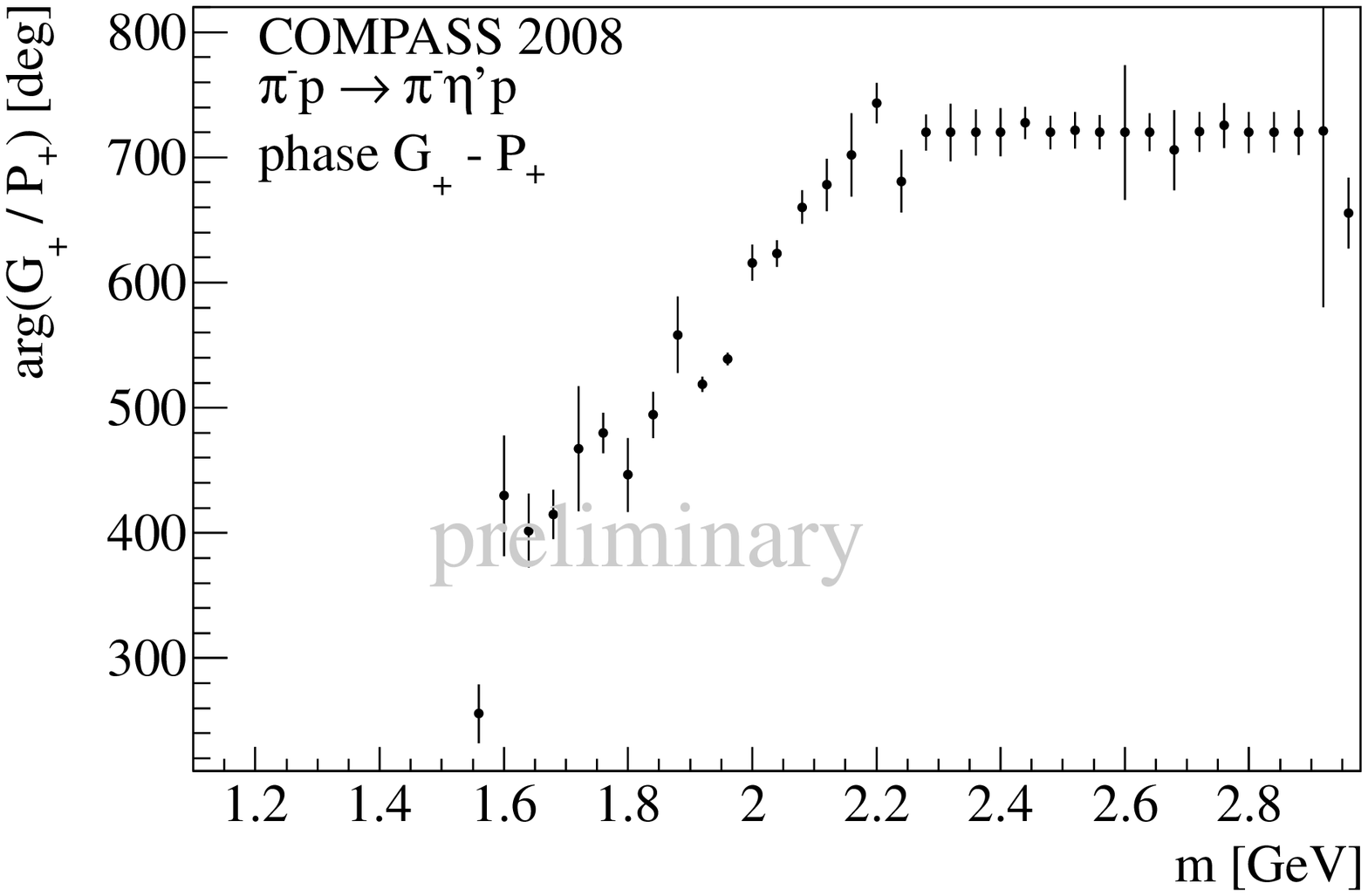}
    \includegraphics[width=0.3\textwidth]{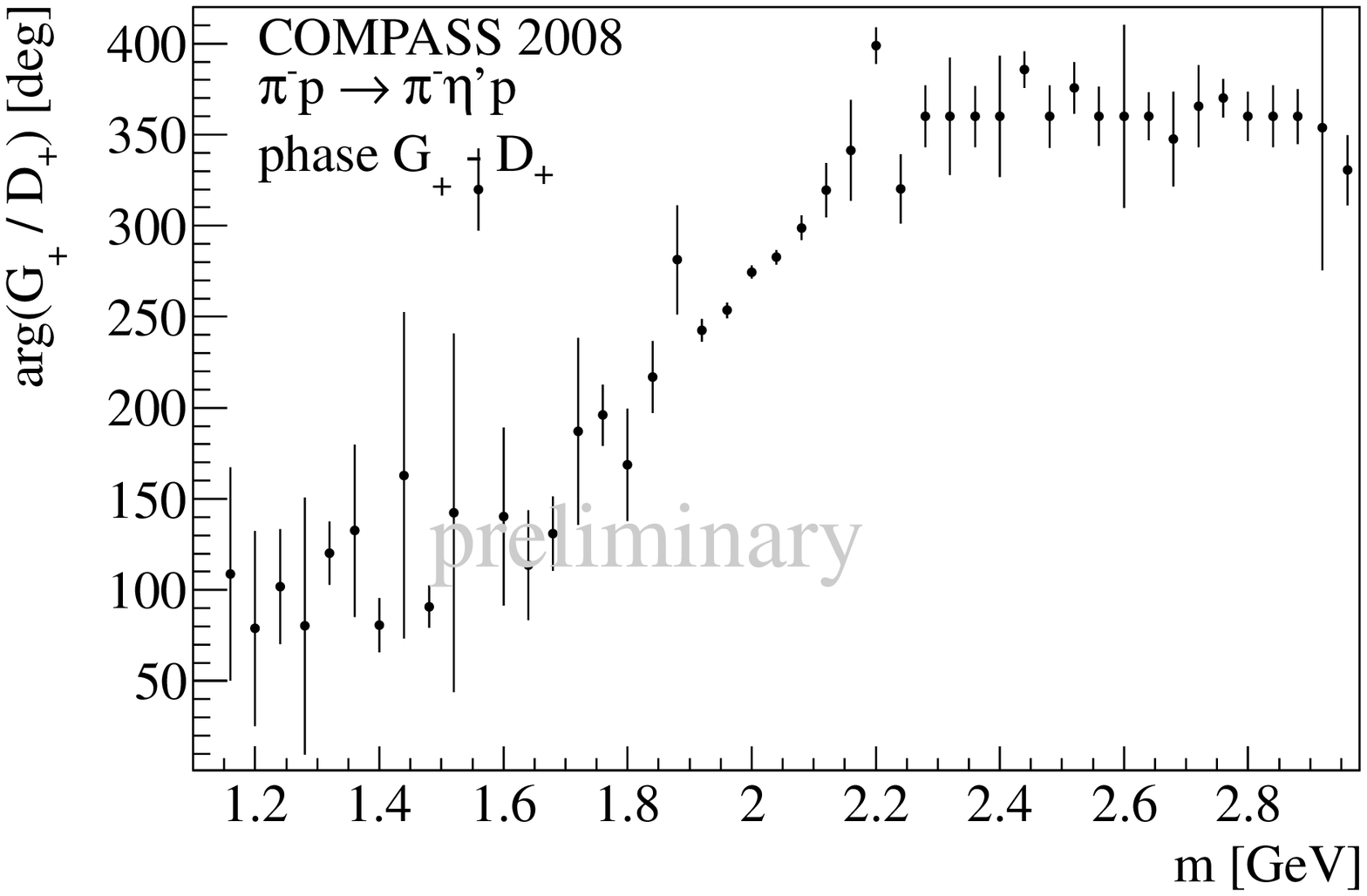}
    \caption{Relative phases of the positive-reflectivity waves.  From
      left to right in obvious notation: $D_+-P_+$, $G_+-P_+$, $G_+-D_+$.}
    \label{fig:phases}
  \end{center}
\end{figure}

\acknowledgements{%
  We acknowledge financial support by the German Bundesministerium
  f\"ur Bildung und Forschung (BMBF), by the
  Maier-Leibnitz-Laboratorium der LMU und TU M\"unchen, and by the DFG
  cluster of excellence ``Origin and Structure of the Universe''.  }


\providecommand{\href}[2]{#2}\begingroup\raggedright\endgroup

%

}  


\end{document}